\documentstyle[times,pramana,epsf,floats]{ias}
\begin{document}
\mark{{Gravitational collapse...}{Pankaj S. Joshi}}
\title{Gravitational Collapse: The Story so far}

\author{Pankaj S. Joshi}
\address{Tata Institute of Fundamental Research, Homi Bhabha Road,
Colaba, Mumbai 400005}
\pacs{2.0}
\abstract{An outstanding problem in gravitation theory and relativistic 
astrophysics today is to understand the final outcome of an endless
gravitational collapse. Such a continual collapse would take place when 
stars more massive than few times the mass of the sun collapse 
under their own gravity on exhausting their nuclear fuel. According to
the general theory of relativity, this results either in a black hole, 
or a naked singularity- which can communicate with faraway observers in the
universe. While black holes are (almost) being detected and are 
increasingly used to model high energy astrophysical phenomena, naked
singularities have turned into a topic of active discussion, aimed at 
understanding their structure and implications. Recent developments here
are reviewed, indicating future directions.}

\maketitle
\section{Introduction}

What is the final outcome of the continual gravitational collapse of
a massive star which has exhausted its nuclear fuel? While stars not very
massive could stabilize as white dwarfs or neutron stars, any stellar core
more massive than about five solar masses must collapse endlessly according
to our present physical understanding. The question of final fate of such
an endless collapse is of central importance in gravitation theory
and astrophysics today. The theory to use to examine this question related 
to strong gravity fields is the general theory of relativity, which should 
be valid till the quantum gravity length scales of about $10^{-33}$cms, and 
which would be a low energy limit of any reasonable quantum gravity theory.  

In early seventies, the singularity theorems in general relativity gave 
some partial hints to an answer to the above question. Under a reasonable 
set of physical assumptions, such as causality and positivity of 
energy density, these theorems showed that closed trapped surfaces, which 
develop in gravitational collapse, give rise to spacetime singularities. 
Such singularities signal the onset of a phase of extreme strong gravity 
regions where the quantum effects should start getting important. 
It is only in such strong gravity fields near singularities where both 
general relativity and quantum gravity come into their own, and one 
may have an opportunity to test the effects of quantum gravity. The 
limitation of the singularity theorems, however, was that they only 
predicted the existence of singularities in collapse and cosmology, but 
did not give any information on their {\it physical nature} (e.g.  how 
fast the densities and curvatures grow there), or the {\it causal 
structure} (e.g. can they communicate with faraway observers in the 
universe).  

These questions, which are clearly vital to understanding the final
fate of massive collapsing clouds, can be answered only by means of a 
detailed study of gravitational collapse phenomena in gravitation 
theory\cite{rev}.
The classical spacetime singularities should be smeared out by quantum
gravity, and what would really result from such an endless collapse is 
an extreme strong gravity region, with extreme values of physical parameters
such as densities and curvatures, confined to an extraordinarily tiny 
region of space. If the event horizons of gravity already start developing 
at an earlier phase during such a collapse, the collapsing star and the 
eventual fireball as described above gets hidden within the horizon, 
disappearing from the purview of the outside observers in the universe 
forever. Then we have the formation of a {\it black hole} in the universe
as a result of the gravitational collapse. On the other hand, if the 
formation of event horizon gets delayed sufficiently during the collapse, 
the result is the development of a {\it naked singularity}, or a 
{\it visible fire ball}, which can possibly send out massive radiations 
to faraway observers from near such strong gravity regions.

A detailed study of gravitational collapse phenomena from such a 
perspective has been conducted within the context
of classical gravity\cite{psjbook}. The generic conclusions emerging
from these studies are striking: {\it While the collapse always produces 
curvature generated fireballs characterized by diverging curvatures and
densities, trapped surfaces may not develop early enough to always 
shield this process from an outside observer}. Specifically, as we 
shall show in the next section, depending on the nature of the initial 
data from which the collapse develops, either a black hole or a naked 
singularity results as the final outcome of the collapse. We then
discuss several implications and generalizations of these results,
giving an idea of the recent developments and future directions 
in this field.

\section{Spherically symmetric collapse}

Spherically symmetric collapse has been investigated from such a 
perspective in detail. The first studies examining the dynamical 
evolutions of collapsing matter clouds were due to Oppenheimer and Snyder, 
and Dutt\cite{osd}. As is well-known now, such a collapse of a 
homogeneous dust ball (the density and pressures given by 
$\rho=\rho(t), p=0$) gives rise to a black hole, where the extreme
density regions are necessarily hidden from the faraway observer by the
event horizon, which starts forming much earlier than the epoch of
formation of singularity.

What is the outcome of the collapse, however, when the density is 
allowed to be inhomogeneous, which is a physically more realistic 
situation? 
The collapse of spherically symmetric inhomogeneous dust has been 
studied in detail, and we now know that the outcome is generically
either a black hole or a naked singularity, depending on the 
nature of the initial data 
from which the collapse develops\cite{dust}.

One may, however, consider dust as somewhat unrealistic form of matter,
especially towards the end stages of collapse, when pressures should be
important. From such a perspective, the gravitational collapse of 
perfect fluids, and other more general forms of matter, has been 
studied analytically\cite{analy} and also numerically\cite{numer}. The 
conclusions remain essentially the same,
namely, both black holes and naked singularities do develop as end state
of gravitational collapse.

In the following, using the general treatment given by Joshi and Dwivedi
\cite{jd99} we show that given an arbitrary regular distribution of 
a general matter field at the initial epoch, there always exists an 
evolution from this initial data which would result either in a black 
hole or a naked singularity, depending on the allowed choice of free 
functions available in the solution. It follows that either of these 
objects result depending on the nature of the regular initial data 
from which the collapse evolves. Again the usual energy conditions ensuring 
the positivity of energy density and other regularity conditions will be
satisfied. This method also generates wide new families 
of black hole solutions resulting from spherically symmetric collapse,
without requiring the cosmic censorship assumption.

We consider here general type I matter fields\cite{he}, which 
include most of the physically important forms of matter such as dust, 
perfect fluids, massless scalar fields and so on. In fact, almost 
all observed forms of matter and equations of state are included in
this general class.  Our purpose is
to analyze the collapse with a given initial data set such as
the state of matter and the velocities of the
spherical shells at the onset of collapse for a compact object, 
in order to determine the possibilities of this configuration evolving 
into either a black hole or a naked singularity.
So we consider the gravitational collapse of a matter cloud 
that evolves from a regular initial data defined on an initial 
spacelike surface. The energy-momentum tensor has a compact support 
on this initial surface where all the physical quantities such as 
densities and pressures are regular and finite.

Such a matter field, in a general coordinate system, can be expressed as
\begin{equation}
T^{ab}=\lambda_1 E^a_1E^b_1 +\lambda_2 E^a_2E^b_2
+\lambda_3 E^a_3E^b_3 +\lambda_4 E^a_4E^b_4
\end{equation}
where (${E_1,E_2,E_3,E_4}$) is an orthonormal basis with ${E_4}$ 
and $(E_1,E_2,E_3)$ are timelike and spacelike eigenvectors 
respectively, and $\lambda_{i}$s ($ i=1,2,3,4$) are the eigenvalues.
For such a spherically symmetric matter distribution
we can choose coordinates $(x^i=t,r,\theta,\phi)$ adopted to this
orthonormal frame, and the metric is written as,
\begin{equation}
ds^2=-e^{2\nu}dt^2+e^{2\psi}dr^2+R^2d\Omega^2
\end{equation}
where $d\Omega^2= d\theta^2+ \sin^2\theta d\phi^2$ is the line element
on a two-sphere. 
Here $\nu,\psi$ and $R$ are functions of $t$ and $r$, and the stress-energy
tensor $T^a_b$  as given by equation (1) 
has only diagonal components in this coordinate system (i.e. we are
using a comoving coordinate system), given by 
\begin{equation}
T^t_t=-\rho,\quad T^r_r=p_r,\quad T^{\theta}_{\theta}=p_{\theta}=
T^{\phi}_{\phi},\quad T^t_r=T^r_t=0
\end{equation}
The quantities 
$\rho, p_r$,  and $p_{\theta}$ are the eigenvalues of $T^a_b$ and are
interpreted as the density, radial pressure, and tangential stresses 
respectively for the cloud. We take the matter fields to satisfy 
the weak energy condition,
i.e. the energy density as measured by any local observer must be
non-negative, and so for any timelike vector $V^a$ we must have 
\begin{equation}
T_{ab}V^aV^b\ge 0
\end{equation}
which amounts to
\begin{equation}
\rho\ge0,\quad \rho+p_r\ge0,\quad \rho+p_{\theta}\ge 0
\end{equation}

From the point of view of the dynamical evolution of the initial
data at an epoch of time from which the collapse commences, we have
a total of six arbitrary functions of $r$, namely,
\begin{equation}
\nu(t_i,r)=\nu_o(r),\quad \psi(t_i,r)=\psi_o(r),\quad R(t_i,r)=R_o(r),
\end{equation}
\begin{equation}
\rho(t_i,r)=\rho_o(r),\quad p_r(t_i,r)=p_{r_o}(r),\quad p_{\theta}(t_i,r)
=p_{\theta_o}(r)
\end{equation}
These functions constituting the initial data are to be specified at 
an initial surface at an initial epoch $t=t_i$.
The dynamical evolution of this initial data is determined by the
Einstein equations, and for the metric (2) these are given by,
\begin{equation}
T^t_t=-\rho=-{F'\over k_oR^2R'},\quad
T^r_r=p_r=-{\dot F\over k_oR^2 \dot R}
\end{equation}
\begin{equation}
\nu'(\rho +p_r)=
2(p_{\theta}-p_r){R'\over R}-p_r'
\end{equation}
\begin{equation}
-2\dot R'+R'{\dot G\over G}+\dot R {H'\over H}=0
\end{equation}
\begin{equation}
G-H=1-{F\over R}
\end{equation}
where $(\dot{ })$ and $( ' )$ represent partial derivatives with respect to
$t$ and $r$ respectively, $F=F(t,r)$ is an arbitrary function of $t$ and $r$, 
and we put
\begin{equation}
G=G(t,r)=e^{-2\psi}(R')^2,\quad H=H(t,r)=e^{-2\nu}\dot R^2
\end{equation}
The function $F(t,r)$ here is treated 
as the mass function for the cloud, with $F\ge 0$. In order to preserve the
regularity of the initial data at $t=t_i$, we must have $F(t_i,0)=0$, that
is, the mass function vanishes at the center of the cloud.

The initial data represented by the functions $\nu_o,\psi_o,\rho_o,
p_{r_o},p_{\theta_o}$ and $R_o$ are not all independent. 
The equation (9), when evaluated on the initial surface, gives the
relationship
\begin{equation}
\nu_o'(\rho_o +p_{r_o})=
2(p_{\theta_o}-p_{r_o}){R'_o\over R}-p_{r_o}'
\end{equation}
Furthermore, there is a coordinate freedom left in the choice of 
the scaling of the coordinate $r$, which can be used to reduce the 
number of independent initial data to four. Thus 
there are only four independent arbitrary functions of $r$  
constituting the initial data.
Evolution of this data is governed by the field equations,
and we have in all five equations with seven unknowns,
namely $\rho,p_r,p_{\theta},\nu,\psi,R$ and $F$, giving us freedom of choice
of two functions. Selection of these two free functions, subject to the given
initial data and the weak energy condition above, 
determines the matter distribution and the metric of the 
spacetime and thus leads to a particular evolution for the initial data. 
We need to ensure the regularity of the initial data,
which would be the case if the curvatures
and the initial data describing the matter (the initial density and 
pressures) are all finite. For curvatures to be finite
one must have a bounded Kretchmann scalar $K=R^{abcd}R_{abcd}$.
A singularity will appear on the initial surface 
if either the density $\rho$, or 
one of the pressures become unbounded at any point on the initial 
surface, or $(F/R^3)\rightarrow \infty$ at any point. 
We require that on the initial surface $t=t_i$, the density and pressures 
are finite and bounded. 
We also have 
\begin{equation}
F(t_i,r)=\int{\rho_o(R_o)R^2_odR_o}
\end{equation}
and hence the spacetime is singularity free initially in the sense that
the Kretchmann scalar, density and pressures are all finite. But, as the
collapse evolves, a singularity could develop at a later time whenever 
either of the density or one of the pressures become unbounded. 
We shall consider below such specific evolutions of the initial data
which model a gravitationally collapsing matter cloud.

Using the coordinate freedom left in rescaling the radial 
coordinate, we rescale $r$ such that
\begin{equation}
R(t_i,r)=R_o(r)=r
\end{equation}
The physical area radius $R$ then monotonically increases with the 
coordinate $r$, and there are no shell-crossings on the initial surface,
with $R'=1$. Since we are considering gravitational collapse,
we also have $\dot R<0$. 
The initial data for a collapsing matter cloud is given in terms
of the initial densities and pressures describing the initial state of matter 
at the onset of collapse. These are $\rho_o,p_{r_o}, p_{\theta_o}$,
and the function $\psi_o$, related to the initial velocity of 
the collapsing shells. Next, physically reasonable
matter forms may satisfy an energy condition ensuring the positivity
of mass-energy densities. Therefore, at the onset of the collapse, 
all the initial data sets specifying the density and pressures profiles 
of the cloud satisfy an energy condition, and the same holds during
the later evolution of collapse. Satisfying the weak energy condition 
implies for the initial data,
\begin{equation}
\rho_o(r) \ge 0,\quad \rho_o+p_{r_o}\ge 0,\quad \rho_o(r)+
p_{\theta_o}(r)\ge 0,
\end{equation}
and the same holds at all later epochs of collapse.

Because at the initial epoch $t=t_i$, $\nu_0 \ne \pm\infty$, we have 
from Einstein 
equations some restrictions on the arbitrariness of the choice 
of the functions $\rho_o,p_{r_o},p_{\theta_o}$. We have, for example, 
\begin{equation}
[p_{\theta_o}-p_{r_o}]_{r=0}=0 
\end{equation}
For the sake of physical reasonableness, we require the center $r=0$ to be 
the regular center for the cloud, which means $R(t,0)=0$.  
Also, one would like to have the initial density $\rho_o(0)>0$ 
at the center $r=0$. This implies that we have
\begin{equation}
\rho_o(0) +p_{r_o}(0)>0,\quad  \nu_o(r)=r^2h(r)
\end{equation}
where $h(r)$ is at least a $C^1$ function of $r$ for $r=0$, and at least
a $C^2$ function for $r>0$.
This means that the pressure gradients 
vanish at the center $r=0$, basically meaning that the forces vanish 
at the center. Here we consider this scenario for the 
sake of physical reasonableness, however, it is possible to give a more 
general formalism independent of requirements such as above.

Actually, it would be reasonable to require the pressures to be positive 
at the onset of the collapse, since for astrophysical bodies physically 
we would prefer the pressures rather than tensions. 
Further more, to make the scenario physically more appealing, we may 
require the density to be decreasing as we move away from the center $r=0$.
In that case, for any reasonable equations of state such as 
$p=k\rho, 0<k<1,$ (a perfect fluid), or $p=k\rho^\gamma$, the pressure also
will decrease away from the center together with the decreasing density.
This may typically be the case in the massive bodies such as stars and such
other astrophysical systems. Then as such the energy conditions  
could impose restrictions on the maximum size of the matter cloud with 
such an initial density and pressure distribution. Additionally, for a
compact collapsing star, at the onset of the collapse the radial stress 
should vanish at some boundary $r=r_b$ i.e. $p_{r_o}(r_b)=0$.

For the given initial data set of density and pressures there may be 
infinitely many collapse evolutions possible. However, we 
do not intend to
find a particular solution of the field equations prescribing a particular
evolution of any given initial data. We investigate here, within 
the framework of general relativity and the regular initial data to
begin with, when physically allowed reasonable evolutions
develop into naked or covered singularities. The answer to this question
would have been simple, if we had at our disposal both an exact 
closed equation
of state describing the state of collapsing matter, and an exact solution
of the field equation. However, both of these are little understood in
relativity in highly dense regions. In dust models
the initial data describing the matter consists of only the initial density 
distribution. Dust models could be criticized due to the 
vanishing pressures,
and it is possible that if pressures are present than the conclusions 
regarding the final fate of collapse could
be different. We therefore look for an evolution of an arbitrary initial
data set consisting of both the density and non-zero pressures, and which
would reduce to dust if initial pressures are vanishing. In
dust models we have that the mass function $F=F(r)$ is time independent,
and $\nu=0$. We
therefore consider an ansatz which is a simple extension of dust, and
such that one can still incorporate initial non-vanishing pressures.

Consider the gravitational collapse of a matter cloud with a 
general initial data as prescribed above, and the functions
$F$ and $\nu$ given as below,
\begin{equation}
\nu=c(t)+\nu_o(R),\quad F=f(r)+F_o(R)
\end{equation}
From the Einstein equations, the evolution of collapse 
is then described by the equations,
\begin{equation}
\nu_o(R)=R^2g(R)=\int_0^R {\left({2p_{\theta_o}-2p_{r_o}\over r(
\rho_o +p_{r_o})}-{p_{r_o}'\over \rho_o +p_{r_o}}\right)dr}
\end{equation}
\begin{equation}
G=b(r)e^{2\nu_o}
\end{equation}
\begin{equation}
\sqrt{R}\dot R=-a(t)e^{\nu_o}\sqrt{b(r) Re^{\nu_o}-R+f+F_o}
\end{equation}
\begin{equation}
\rho={f'\over R^2R'}+{F_o,_R\over R^2}, \quad p_r=-{F_o,_R\over R^2}
\end{equation}
\begin{equation}
2p_{\theta}=R\nu,_R(\rho +p_r)+2p_r+Rp_r,_R
\end{equation}
\begin{equation}
F_o(R)=-\int_0^R{r^2p_{r_o}dr}\equiv -R^3{\it p}(R)
\end{equation}
\begin{equation} 
f(r)=\int_0^r{r^2(\rho_o+p_{r_o})dr}\equiv
r^3\epsilon(r)+r^3p(r)
\end{equation}

Here $F(t_i,r)=2r^3\epsilon(r)$. The quantities $\epsilon(r)$ and 
$p(r)$ are to be treated 
as the average mass and pressure densities of the cloud, and are 
decreasing functions of $r$. Since $p_{r_o}$ is a positive function on
the initial surface, it follows that $F_o<0$ throughout the spacetime, 
and as such the radial pressure is non-negative throughout the spacetime.
The arbitrary function $b(r)$ characterizes the velocity of the spherical 
shells at the initial time $t=t_i$. We are dealing with the collapse 
situation with $\dot R <0$, therefore the arbitrary function $a(t)>0$.

We note that in the above model for evolution of the
initial data set as described by the above equations, 
if the initial pressures 
vanish, i.e. $p_{r_0}(r)=p_{\theta _o}(r)=0$, we then have as the solution
\begin{equation}
ds^2=-e^{a(T)}dT^2+{R_T'^2\over 1+E(r)}dr^2+R_T^2d\Omega^2
\end{equation}
\begin{equation}
\nu= p_r=p_{\theta}=0\quad G=b(r)\equiv 1+E(r)\quad \dot R_T=-\sqrt{
E+{F_T(r)\over
R_T}}
\end{equation}
where we have changed the notation for $t\to T$ and $R\to R_T$ 
in order to distinguish the above solution, 
which is actually the Tolman-Bondi-Lemaitre solution for a dust cloud.
Thus, the models here may also be viewed as directly generalizing
the Tolman-Bondi-Lemaitre models to include both the radial and
tangential pressures in order to investigate the role of initial
data towards the final fate of collapse.

Another important point that one has to consider is the matching 
of the gravitational collapse model above with a suitable exterior, 
which is either an asymptotically 
flat region, or a cosmological background. For details on this, as well as 
a discussion of energy conditions during the evolution of collapse,
we refer to \cite{jd99}.

Since $r=0$ is the regular center of the cloud, meaning $R(t,0)=0$,
it follows from equation (22) that 
\begin{equation}
\sqrt{v}\dot v=-a(t)e^{\nu_o}\sqrt{v^3(h(R)b(r)-{\it p}(R))+ b_o(r)v+
\epsilon(r)+{\it p}(r)}
\end{equation}
where the arbitrary function $b(r)=1+r^2b_o(r)$, such that $b_o(r)$ is 
at least a $C^1$ function of $r$ for $r=0$, and a $C^2$ function for $r>0$, 
and we have introduced
\begin{equation}
R=rv(t,r),\quad v(t_i,r)=1
\end{equation}
\begin{equation}
h(R)=h(rv)={e^{2r^2v^2g(rv)}-1\over r^2v^2} ={e^{2\nu_0}-1\over R^2}
\end{equation}
The functions $b_o(r), h(rv), v(t,r), g(rv)$, and $f_o(r)$ are all at least
$C^1$ functions of their arguments. At $t=t_i$ we have 
$v=1$ and since $\dot v <0$, we have $v<1$ throughout the spacetime.

The quantity $R(t,r)\ge 0$ here is the area radius 
in the sense that $4\pi R^2(t,r)$ gives the proper area 
of the mass shells at any given value of the
comoving coordinate $r$ for a given epoch of time. The area of a 
shell at $r=$const. goes to zero when $R(t,r)=0$. In this sense, the curve 
$t=t_s(r)$ such that $R(t_s,r)=0$
describes the singularity in the spacetime where the mass shells are
collapsing to a vanishing volume, with the density and pressures
diverging. This shell-focusing singularity occurs along the curve
$t=t_s(r)$ such that $v(t_s,r)=0$, the Kretchmann scalar diverges
at such points. Using the remaining degree of freedom left
in the scaling of the time coordinate $t$ we could set $a(t)=1$. 
Equation (29) can then be integrated with the initial condition $v(t_i,r)=1$ 
to obtain the function $v(t,r)$.
We get
\begin{equation}
\int_v^1{\sqrt{v}dv\over \sqrt{b_o(r)ve^{3\nu_o}+e^{2\nu_o}(v^3(h(rv)-p(rv))
+\epsilon(r)+{\it p}(r))
}}=t
\end{equation}
where we have chosen for the sake of simplicity $t_i=0$. 
Note that the coordinate $r$ is treated as a constant in the equation.
The time $t=t_s(r)$ 
that corresponds to the occurrence of singularity is then given by,
\begin{equation}
t=t_s(r)=\int_0^1{\sqrt{v}dv\over 
\sqrt{b_o(r)ve^{3\nu_o}+e^{2\nu_o}(v^3(h(rv)-p(rv))
+\epsilon(r)+{\it p}(r))
}}
\end{equation}

In gravitational collapse, a singularity can also occur at 
$R'=0$, which is called a shell-crossing singularity.
But these are singularities of a weaker nature in general, and the spacetime 
can possibly be extended through these using a suitable extension
procedure. However, the comoving coordinate system we use
here may break down and the metric may possibly become degenerate 
at the points where $R'=0$. Our purpose here is to study the 
shell-focusing singularity at $R=0$, which is essentially different and
could be much stronger gravitationally as compared to the shell-crossings 
which are delta-function like singularities, caused by different 
shells crossing each other where the density momentarily blows up. 
Hence, we choose the evolution of the initial data in such a 
manner that any shell-crossings are avoided in the collapse, except
possibly at the singularity. Here we mention that
a similar situation regarding the occurrence of shell-crossings arises 
in Tolman-Bondi-Lemaitre dust collapse models also. However, as
has been pointed out in earlier works, for a given initial density
profile one can always choose appropriate initial velocity of the
dust shells such that during the evolution no shell-crossings are
encountered or visa-versa.

In the present general case also, the same can be achieved 
by means of a suitable choice of the functions involved
as specified below. At a given epoch of time, the functions $\epsilon (r)$, 
$p(r)$, $\nu_o(R)$ and $h(R)$ are at least $C^1$ functions and also 
$\epsilon(r)$ and $p(r)$ are decreasing functions of $r$. Then 
$b_o(r)$ is an arbitrary function representing the initial velocities 
of the collapsing spherical shells. From equation (33) it is clear that 
the singularity time $t_s(r)$ is an explicit
function of the velocity $b_o(r)$, which is a free function,
and one can choose it in such a way that $t_s(r)$ is
an increasing function of $r$, i.e. $dt_s/dr >0$. The exact nature of such
velocity functions $b_o(r)$ for which $t_s(r)$ is an
increasing function of $r$ depends upon the exact behavior of initial density
and pressures within the cloud. For example, for a matter
cloud initially satisfying an equation of state of the type $p=a\rho^{\gamma}$
one of the many possibilities is the function $b_o(r)>0$ such
that $b_o'(r)<0$ and is less than a certain minimum for $r_b\ge r\ge 0$. 
For all such functions 
$b_o(r)$, therefore, the singularity curve $t=t_s(r)$ is an increasing curve 
for all allowed values of coordinate $r$, and 
the successive spherical shells within the cloud
collapse to singularity successively, and shell-crossings do not
occur. Thus $R'=v+rv'>0$ (note that $R'=1$ initially) 
throughout the spacetime.
Furthermore, if $[r^2b_o]'\ge 0$ then  $\sqrt{v}R'\le 1$ within 
the cloud for $1\ge v\ge 0$.

The central shell-focusing singularity $R=0$ occurs first at $r=0$ 
and the time of occurrence of such a singularity, using the above
equations, is given by,
\begin{equation}
t_{s_o}=t_s(0)=\int_0^1{\sqrt{v}dv\over 
\sqrt{v^3(h_o-p_o)+b_{oo}v+\epsilon_o+p_o}}
\end{equation}
where $h_o=h(0),p_o=p(0), \epsilon_o=\epsilon(0), b_{oo}=b_o(0)$ 
are constants related to the central density and pressures.

In fact, near the center $r=0$ we have
\begin{equation}
t_s(r)=t_{s_o}+rX(0)+ O(r^2)...
\end{equation}
where the function $X=X(0)$ is given by
\begin{equation}
X(0)=\int_0^1{{\sqrt{v}(\epsilon_1+p_1+b_1v-v^4h_1)dv\over 
(v^3(h_o-p_o)+b_{oo}v+\epsilon_o+p_o)^{3/2}}}
\end{equation}
where $\epsilon_1=-\epsilon'(0), p_1=-p'(0), b_1=-b_o'(0), h_1=h,_R(0)$.

For the case we have been considering, where pressures have been 
taken to 
be positive, the central singularity at $r=0$ could be naked, but 
all subsequent 
singularities with $r>0$ are covered as the quantity $F/R\rightarrow \infty$
and the trapped surfaces and the apparent horizon develop prior to the
formation of the singularity. We note that when the pressures are allowed 
to be negative, still subject to the validity of the weak energy 
condition, the other parts of the singularity can be visible in 
principle\cite{cooper}. It thus remains to examine only the nature of the 
central singularity.

Within the collapsing cloud the apparent horizon is given by 
the condition $R/F=1$. It is the boundary of the trapped surface region 
in the spacetime. The behavior
of the apparent horizon curve (which meets the central singularity at
$R=r=0$) near the center essentially determines
the visibility, or otherwise, of the central singularity. For example,
it is known within the context of the Tolman-Bondi-Lemaitre models that
the apparent horizon is either past pointing timelike or null, or it can
be spacelike, as is seen by examining the nature of the induced metric 
on this surface. This is unlike the event horizon which is always 
future pointing null. If the neighborhood of the center gets trapped earlier 
than the singularity, then it is covered, and if that is not the case the
singularity can be naked, with families of nonspacelike trajectories
escaping from it.

To examine the existence or otherwise of such families,
and to examine the nature of the central singularity occurring at 
$R=0, r=0$ in the general class of models being considered here,
let us consider the equation of the outgoing radial null geodesics
which is given by,
\begin{equation}
{dt\over dr}=e^{\psi-\nu}
\end{equation}
The singularity appears at the point $v(t_s,r)=0$, which corresponds to 
$R(t_s,r)=0$. Hence, if there are future directed outgoing radial null 
geodesics, terminating in the past at the singularity, then along these 
trajectories we have $R\to 0$ as $t\to t_s$.
Writing the equation for these radial null geodesics in terms of the 
variables $(u=r^{5/3},R)$ we obtain, 
\begin{equation}
{dR\over du}= {3\over 5}({R\over u}+{\sqrt{v}v'\over {(R/u)}^{1/2}}
){1-{F\over R}\over \sqrt{G}(\sqrt{G}+\sqrt{H})}
\end{equation}
If the null geodesics terminate in the past at the singularity with a 
definite tangent, then at the 
singularity the tangent to the geodesics $dR/du>0$ in the $(u,R)$ plane,
and must have a finite value.
In the case of collapsing matter cloud we are considering, all 
singularities at $r>0$ are
covered since $F/R\rightarrow \infty$, and therefore $dR/du \rightarrow 
-\infty$. So only the singularity at center $r=0$ could be naked. 
As mentioned earlier, for the case when $R'>0$ near the central singularity, 
we have
\begin{equation}
x_o=\lim_{t\to t_s,r\to 0}{R\over u}=\lim_{t\to t_s,r\to 0}{dR\over du}
\Rightarrow x_o^{3/2}={3\over 2}X(0)
\end{equation}
where $X(0)$ is given by equation (36). Because  $X(0)>0$ the singularity is 
at least locally naked. The behavior
of outgoing radial null geodesics in the neighborhood of the singularity
are described by $R=x_ou$ in $(R,u)$ plane and in $(t,r)$ plane it
is given by
\begin{equation}
t-t_s(0)=x_or^{5/3}
\end{equation}

One can also write the equation for these radial null geodesics
in terms of the variables $(t,R)$ in order to see how the area radius 
$R$ grows along these outgoing null geodesics with increasing 
values of time. As mentioned earlier, for the case where $R'>0$ near 
the central singularity, we get
\begin{equation}
X_o=\lim_{t\to t_s,r\to 0}{R\over t-t_s(0)}=
\lim_{t\to t_s,r\to 0}{dR\over dt}= 
=\lim_{t\to t_s,r\to 0}[e^{\nu}{1-{F\over R}\over \sqrt{G}+\sqrt{H}}]
=1
\end{equation}
Again this shows that the singularity is naked at least locally. 
In fact, as pointed out above, it follows that for
$b_o'(0)\ne 0$ the area coordinate behaves as $R=const.\times r^{5/3}$
near the singularity.

The global visibility of such a singularity, which is locally naked
as above, will depend on the overall
behavior of the various functions concerned within the matter cloud and we
shall not go into those details presently. It has been seen, however, from the
study of various examples so far, that once the singularity is locally naked,
one can almost always make it globally visible by a suitable choice of 
allowed functions. Note that in cases where the choice of $b_o(r)$ 
is such that $X(0)<0$ the singularity would be covered.
When $X(0)=0$, one has to consider the next higher order expansion term
which is nonvanishing in the equation (35), and that will then determine the
nature of the singularity by essentially a similar analysis.

\section{Open issues}

It may be fair to conclude, as pointed out above, that  
generically the gravitational collapse of a massive matter cloud would
produce either a black hole or a naked singularity as the final state,
depending on the nature of the initial data developing the collapse.
The latter would essentially consist of densities and velocities profiles,
and the velocities of the collapsing shells. Such models respect the
physical reasonableness requirements such as positivity of energy
densities, and regularity of initial data.

Such a scenario gives rise to several important questions and open
problems, some of which are stated below. While our discussion may 
not be exhaustive, we hope this will point to some
future directions in this area.

\subsection{How to formulate the cosmic censorship conjecture?}

Actually, it was this very important and basic question which 
led me into my studies on gravitational collapse. Since there 
was no rigorous formulation available for this hypothesis, not 
to speak of a proof, what was really needed I thought was a 
detailed and deeper study of the collapse phenomena in gravitation 
theory. It does not appear we are any closer to an 
answer still.

What has happened, however, is we know now that many of the 
possibilities suggested earlier towards formulating this conjecture
do not work. Various past suggestions included, for example, that naked
singularities will not occur when energy conditions are obeyed, 
or even if they
occur they will be gravitationally weak and removable, or no naked
singularities occur when we allow for pressures, or when we use a 
reasonable form of matter and equation of state, or that in realistic 
cases only a zero measure set of photon and particle trajectories 
come out, and so on. As pointed out above, naked singularities do 
occur even when we impose such `reasonable' conditions. The real issue 
then, in the light of our current knowledge of gravitational collapse 
phenomena, is the genericity and stability of these objects in 
gravitation theory. Thus, a formulation I propose is: 

{\it No naked singularities forming in gravitational collapse of 
reasonable matter fields, developing from regular initial conditions, 
can be generic or stable.}  
   
Of course, one would need to define and formulate these concepts 
of `genericity'
and `stability' in gravitation theory in a much more precise manner,
and that is no easy task which may require sophisticated mathematical
tools. We describe some attempts in that direction below.
Then, trying to prove such a version will be the next challenge.
Finally, if cosmic censorship fails in classical gravity, it is 
possible that quantum gravity may provide a hope, restoring some
version of a censorship in the universe.

\subsection{Do naked singularities occur in non-spherical collapse?}

The spherically symmetric collapse has been studied extensively,
as discussed above. It is then important to know if the conclusions
derived in this case 
hold for non-spherical collapse as well. This issue is largely open,
as there are no good models available presently for studies of this kind.

Some indicative studies are available, however, which tell us that
naked singularities are not necessarily ruled out as soon as we go away from
sphericity\cite{nonsph}. Shapiro and Teukolsky studied oblate and prolate
collapsing configurations, and Nakamura and others studied spindle and
cylindrical naked singularities. Also, Barabes and Israel made an
analytical study of non-spherical collapse. The quasi-spherical collapse
models due to Szekeres were studied by Joshi and Krolak, who found 
the nature of naked singularities developing in this case to be
very similar to the dust collapse situation. Clearly, more
remains to be done here, and numerical models may be of 
help while studying non-spherical collapse.

\subsection{Are they stable and generic?}

We characterized here wide new families of black hole and naked 
singularity solutions forming in spherical collapse in terms of the 
evolutions of the initial data for the collapsing object. What we still 
do not know is the actual measure of each of these classes in the 
space of all possible evolutions allowed from a given general and 
arbitrary but physically reasonable initial data. This is related closely 
to the issue of stability of naked singularities. As is well-known, the 
stability in general relativity is a complicated issue because there is 
no well-defined formulation or criteria to test stability. Fast evolving 
numerical codes for core collapse models may possibly provide further 
insights here. All the same, these classes appear to be generically 
arising in the collapse models considered here, at least within spherical 
symmetry, in that they are not an isolated phenomena but belong to a general
family. Because, given any density and pressure profiles for the cloud, there 
exists an evolution which will lead to either a black hole or a naked 
singularity as desired, as the end product of collapse. 
Also, while discussing stability and 
genericity, one has to be careful on the criterion one used to test the 
same, because sometimes a criterion is used which makes black holes 
also unstable while trying to show the instability of naked 
singularities.

Given the complexity of the field equations, if a phenomenon 
occurs so widely in spherical symmetry, it is not unlikely that the 
same would be repeated in more general situations. 
In fact, before the advent of singularity theorems, it was
widely believed that the singularities found in more symmetric situations
will go away once we go to general enough spacetimes.

The massless scalar field collapse has been studies in detail 
from such a perspective of genericity, analytically by Christodoulou
\cite{chris}, and numerically by Choptuik and others\cite{chop}. 
In particular, Christodoulou showed that globally naked singularities 
are non-generic for the case of massless scalar field collapse. In 
numerical studies, the collapse was studied for a generic, smooth,
one-parameter family of initial data. There is a critical value for 
the parameter concerned which produces a critical solution which has
a naked singularity.

An important indicator in this connection is the imploding Vaidya 
model, where the singularity is naked when the mass parameter 
$\lambda\le 1/8$, and a black hole develops for $\lambda >1/8$. The
parameter $\lambda$ represents here the initial data in the form of
the rate of mass loss. Thus, the singularities, both naked and covered,
are stable against the perturbation of the parameter $\lambda$, and the point
$\lambda=1/8$ is the critical point indicating the transition from one
phase to the other (see e.g. \cite{psjbook} for details).

A similar situation arises in the case of dust collapse also where
the perturbation in the initial density or velocity distribution within
a certain domain does not alter the nature of the singularity. 
The analysis here in general has a significance in that the nature of the 
singularity is seen to be stable in a certain sense against the perturbation 
of the initial data. It is possible in this case to consider mathematical
structures on the initial data space to examine stability issues more
rigorously\cite{saraykar}. Further, Iguchi, Nakao, and Harada\cite{iguchi} 
examined the stability of dust collapse models against odd-parity 
perturbations, which correspond to rotational motion of the dust fluid. 
Their results indicate that the naked singularity formation process 
appears to be stable against such metric perturbations.

In fact, it was pointed out recently by Mena, Tavakol, and 
Joshi\cite{mena}, who
considered the fully general class of all possible density profiles at
the initial epoch from which the dust collapse develops, that naked 
singularities may be unstable in such a fully general context. However,
they showed that when one considers the class of {\it physically
motivated} density profiles with density higher at the center, and
decreasing away from the center, then the naked singularity formation
stabilizes. This is a potentially important point to bear in mind 
in general debates regarding the stability and genericity of naked 
singularities in gravitational collapse.

\subsection{What are the basic properties and structure of naked 
singularities?}

In order to have a better insight into problems such as above
related to genericity and stability, it would be important to
understand better the structure and basic properties of naked singularities. 
The essential features that emerge from the study 
of gravitational collapse, as described in the previous section, 
towards the structure of naked singularity are the following: 

(a) There is an {\it existence} of a naked singularity, in terms of 
a range available in the parameter space.

(b) There exists a {\it non-zero measure set of families} of
non-spacelike trajectories- photons as well as particles world lines
coming out, as opposed to isolated trajectories.

(c) This turns out to be a powerfully strong curvature singularity,
where the curvature is evaluated in the limit of approach to the 
singularity.

The relevance of this last point is that in such a case it would
not be possible to extend the spacetime through such a singularity, and
it would be an unavoidable feature of the spacetime. We refer to 
\cite{djdnolan} for further details on aspects of strength of 
singularities. In fact, these features are common to the collapse 
models including radiation collapse, dust, perfect fluids, self-similar 
as well as non-self-similar collapse models.

One gets a better insight into the structure and properties of these 
objects by examining exact models. One such class which has been 
examined in detail is that with non-zero tangential pressures, but 
where the radial pressure vanishes\cite{magli}.
Again both naked singularities and black holes form here but several
interesting properties of naked singularities become clearer.

Much insight into the collapse phenomena has been gained by 
studies of self-similar collapse models. The advantage here is, because
of the geometrical symmetry, a complete integration of the photon
and particle trajectories is possible, and many interesting features 
come out. For an excellent recent review see \cite{carr}.

\subsection{What role do the quantum effects play?}

A question frequently asked is: Are singularities, naked or covered,
relevant at all- quantum gravity must wash them away. But this is 
missing the actual issue. One certainly hopes that in a suitable
quantum gravity theory the singularities will be smeared out. However,
the issue is whether the extreme strong gravity regions formed due to
gravitational collapse are visible to faraway observers or not.
Because collapse certainly proceeds classically till the quantum
gravity starts governing the situation at the scale of Planck length 
or so, that is, till the extreme gravity configurations have developed
due to collapse. And it is the visibility or otherwise of such regions
that one is discussing.

The point is, classical gravity implies existence of strong gravity
regions, where both classical and quantum gravity come into their own.
In fact, as pointed out by Wald (see Ref. [1]), if naked singularities
develop, then in a literal sense we come face-to-face with the laws
of quantum gravity whenever gravitational collapse to such an event
occurs in distant regions of our universe. Thus, collapse phenomena
may provide us with a possibility of actually testing quantum gravity 
laws. 

From such a perspective, many studies have been conducted on quantum
effects near naked singularities\cite{quantum}. In particular, Vaz and
Witten worked out the spectrum of quantum radiation from a naked singularity,
in analogy to the Hawking radiation from black holes.
It is possible that quantum effects near the naked singularities
may help us restore some kind of a quantum cosmic censorship, or these
quantum effects could give rise to interesting signatures for naked
singularities.

\subsection{What possible astrophysical implications they may have?}

In the case of occurrence of naked singularity configurations 
developing in gravitational collapse, the emissions of light or particles 
from the ultra-dense regions, i.e. the fireballs, are possible to an 
outside observer in the universe. Would this have observational
consequences? From such a perspective, several works have examined the 
possible astrophysical implications of what happens when collapse of 
a massive star results into a naked singularity, rather than a black hole.

It was examined recently if a naked singularity could be a good 
candidate for a strong source of gravity waves\cite{iguchi}. The 
frequency range at which naked singularities may radiate gravity waves was
estimated by Thorne\cite{kip}. Various observational possibilities
to detect cosmic censorship violations have been suggested recently
by Krolak\cite{krolak}.

Both classical as well as quantum effects in the vicinity of such 
a visible fireball may combine to produce observable signatures for
a faraway observer in the universe. Such
a possibility was explored recently by Joshi, Dadhich, and 
Maartens\cite{roy} in connection with the gamma rays bursts which
remain one of the most intriguing puzzles in astronomy. While collapse
always produces the fireball with diverging curvatures and densities,
late formation of trapped surfaces may allow this mostly radiation 
dominated fireball to expand and create shocks in the 
surrounding medium. In this sense, such collapse generated fireballs
could provide natural candidates for a central engine required for 
the production of gamma rays bursts.

\section{Conclusion}

It appears from considerations such as above that the 
occurrence of singularities, naked or otherwise, is inherent in the 
theory of general relativity, and a distinction between these cases 
may not be possible through general relativity alone.

In fact, the investigations on gravitational collapse phenomena in 
gravitation theory seem to have generated by now a somewhat general 
consensus, that both black holes and naked singularities do develop 
as a result of continual gravitational collapse. The basic question 
remaining is that regarding the genericity and stability of naked 
singularities, whenever they arise in a realistic collapse, as we 
discussed here. These are not, however, the questions easy to conclude 
as there is no standard and unique definition of genericity and 
stability available in gravitation theory.

Under the situation, while efforts are continuing 
to develop such concepts in more precise and better manner, naturally many 
studies have also attempted a better understanding of the nature and
structure of naked singularities, and have tried to investigate their 
astrophysical implications, if any. Such a scenario has turned this  
into a field of quite an active discussion and interest, even to the 
extent of attracting articles in the popular press\cite{nyt}.  

\vfil

{\it Acknowledgements}

It is my pleasure to acknowledge interesting comments and useful 
discussions from Jiri Bicak, Naresh Dadhich, Roy Maartens, 
Sunil Maharaj, Ramesh Narayan, J. V. Narlikar, 
N. Panchapakesan, Reza Tavakol, A. K. Raychaudhuri, P. C. Vaidya, and 
several other friends at ICGC2000.


\begin{thebibliography}{00}
\bibitem{rev} For some recent reviews see e.g. R. M. Wald, gr-qc/9710068; 
R. Penrose, in {\it Black holes and relativistic stars}, ed. R. M. Wald, 
University of Chicago press (1998); P. S. Joshi, in {\it Singularities,
black holes and cosmic censorship}, IUCAA Publications, Pune (1997)
(gr-qc/9702036); A. Krolak, gr-qc/9910108; T. P. Singh, gr-qc/9805066; 
S. Jhingan and G. Magli, gr-qc/9903103.

\bibitem{psjbook} P. S. Joshi, {\it `Global aspects in gravitation and 
cosmology'}, Clarendon Press, OUP, Oxford (1993). 

\bibitem{osd}  J. R. Oppenheimer and H. Snyder, Phys. Rev. {\bf56}, 
p.455 (1939); B. Datt, Z. Physik {\bf108}, p.314 (1938).

\bibitem{dust} D. M. Eardley and L. Smarr, Phys. Rev. D {\bf 19}, (1979) 
p.2239; D. Christodoulou, Commun. Math. Phys. {\bf 93} (1984) p. 171; 
R. P. A. C. Newman, Class. Quantum Grav. {\bf 3} (1986) p.527; B. Waugh 
and K. Lake, Phys. Rev. D {\bf 38} (1988) p.1315; I. H. Dwivedi and P. S. 
Joshi, Class. Quant.Grav., 9, L69 (1992); P. S. Joshi and I.H. Dwivedi, 
Phys. Rev. {\bf D47}, p.5357 (1993); P. S. Joshi and T. P. Singh,
Phys. Rev. D{\bf51}, p.6778 (1995); T. P. Singh and P. S. Joshi, Class.
Quantum Grav. {\bf13}, p.559 (1996); I. H. Dwivedi and P. S. Joshi, 
Class. Quant. Grav. 14, p.1223 (1997); S. Jhingan and P. S. Joshi, Ann. of 
Isr. Phys. Soc., {\bf13}, p.357 (1998). 


\bibitem{analy} P. S. Joshi and I. H. Dwivedi, Commun. Math. Phys. 146, 
p.333 (1992); Lett. Math. Phys. 27, p.235 (1993); Commun. Math. Phys. 
166, p.117 (1994); Class. Quantum Grav. {\bf16}, p.41 (1999); K. Lake, 
Phys. Rev. Lett. 68, p.3129 (1992). 


\bibitem{numer} A. Ori and T. Piran, Phys. Rev. Lett. {\bf59}, p.2137
(1987); ibid., Phys. Rev. D 42, p.1068 (1990); T. Harada, Phys. Rev.
D{\bf58}, p.104015 (1998).


\bibitem{jd99} P. S. Joshi and I. H. Dwivedi, Class. Quantum Grav. {\bf16},
p.41 (1999).

\bibitem{he} S. W. Hawking and G. F. R. Ellis, {\it The large scale
structure of space-time}, Cambridge University Press, Cambridge (1973).

\bibitem{cooper} Cooperstock F., Jhingan S., Joshi P. S. and Singh T. P.,
Class. Quantum Grav. {\bf14}, p.2195 (1997).

\bibitem{nonsph} S. L. Shapiro and S. A. Teukolsky, Phys. Rev. Lett. 
{\bf66}, p.994 (1991); ibid., Phys. Rev. D{\bf45}, p.2006 (1992); 
P. S. Joshi and A. Krolak, Class. Quant. Grav. 13, p. 3069 (1996); 
T. Nakamura, M. shibata and K. Nakao, Prog. Theor. Phys. {\bf89}, 
p.821(1993).

\bibitem{chris} D. Christodoulou, Ann. of Math. {\bf149}, p.183 (1999);
{\bf140}, p.607 (1994); M. Roberts, Gen. Relat. Grav. {\bf21}, p.907 (1989).

\bibitem{chop} M. W. Choptuik, Phys. Rev. Lett. {\bf70}, p.9 (1993);
see C. Gundlach, gr-qc/0001046, and references therein.

\bibitem{saraykar} R. V. Saraykar and S. H. Ghate, Class. Quantum Grav. 
{\bf16}, p.281 (1999).

\bibitem{iguchi} H. Iguchi, K. Nakao and T. Harada, Phys. Rev. D{\bf57},
p.7262 (1998);  H. Iguchi, T. Harada and K. Nakao, Prog. Theor. Phys.,
{\bf101}, p.1235 (1999); ibid., {\bf103}, p.53 (2000).


\bibitem{mena} F. Mena, R. K. Tavakol and P. S. Joshi, gr-qc/0002062.

\bibitem{djdnolan} S. S. Deshingkar, P. S. Joshi and I. H. Dwivedi, 
Phys. Rev D{\it59}, p.044018 (1999); B. C. Nolan, gr-qc/0001026.

\bibitem{magli} G. Magli, Class. Quant. Grav. {\bf14}, p.1937 (1997); 
G. Magli, gr-qc/9711082; T. Harada, K. Nakao and H. Iguchi, Class.
Quantum Grav. {\bf16}, p.2785 (1999); S. Jhingan and G. Magli, gr-qc/9902041; 
S. Barve, T. P. Singh and  L. Witten, Gen. Relat. Grav. {\bf32}, 
p.697 (2000).


\bibitem{carr} B. J. Carr and A. Coley, Class. Quantum Grav. {\bf16},
R31 (1999).

\bibitem{quantum} L. Ford and L. Parker, Phys. Rev. {\bf D17} (1978) 1485;
W. A. Hiscock, L. G. Williams and D. M. Eardley, Phys. Rev. {\bf D26} (1982) 
751; C. Vaz and L. Witten, Phys. Letts. {\bf B325} (1994) 27; 
Class. Quant. Grav. {\bf 12} (1995) 1; {\it ibid.} {\bf 13} (1996) L59; 
Nucl. Phys. {\bf B487} (1997) 409; gr-qc/9804001;
S. Barve, T. P. Singh, C. Vaz and L. Witten, Nucl. Phys. B {\bf532}, 
p.361 (1998);  T. Harada, H. Iguchi and K. Nakao, Phys. Rev. D{\bf61},
p.101502 (2000); ibid., gr-qc/0005114.

\bibitem{kip} K. S. Thorne, in {\it Relativistic Astrophysics},
eds. B. J. T. Jones and D. Markovic, Cambridge University Press (1997).

\bibitem{krolak}  A. Krolak, gr-qc/9910108.

\bibitem{roy} P. S. Joshi, N. Dadhich, and R. Maartens, (gr-qc/0005080) 
Mod. Phys. Lett. A, {\bf15}, p.991 (2000); S. K. Chakravarti and 
P. S. Joshi, (hep-th/9208060) Int. J. Mod. Phys. D{\bf3}, 647 (1994); 
T. P. Singh, gr-qc/9805062; E. Witten, in {\it Quantum teory and beyond},
eds. F. Mansouri and J. Sciano (1992).

\bibitem{nyt} M. W. Browne, New York Times, Feb 12, 1997.



\end{thebibliography}
\end{document}